# Visualization of airborne ultrasound field using thermal images

Ryoya Onishi[1]*, Takaaki Kamigaki[1], Shun Suzuki[1], Tao Morisaki[1], Masahiro Fujiwara[1], Yasutoshi Makino[1], Hiroyuki Shinoda[1]

[1]Department of Complexity Science and Engineering, Graduate School of Frontier Sciences, The University of Tokyo; Kashiwa-shi, Chiba-ken, 277-8561, Japan.
*Corresponding author. Email: onishi@hapis.k.u-tokyo.ac.jp

**ABSTRACT**
Thermoacoustics, the interaction between heat and acoustic waves, is used in a wide range of advanced technological applications. Conversion of ultrasound waves into heat is employed in medical applications, including tumor ablation. However, thermal responses to airborne ultrasounds have not been extensively studied. Herein, we show that thermal responses to high-intensity airborne ultrasounds above 1000 Pa can be used for acoustic measurements. Thermal images on a mesh screen enabled the real-time visualization of three-dimensional acoustic fields. Based on the temperature distribution on a surface that reflects ultrasound waves, it was possible to determine the ultrasound focus position with a resolution much finer than the wavelength. Our results demonstrate that thermography observations based on ultrasound-temperature conversion can be used to visualize acoustic fields, which are conventionally difficult to observe. Our findings will advance technologies that use strong airborne ultrasound, such as mid-air haptics and acoustic levitation, and provide new scientific and technical tools for detecting surface properties and composition, including cells and fragile materials.

The practical use of ultrasound in liquids or solids has attracted significant attention[1]–[7]. Specifically, the application of focused ultrasound for the treatment of tumors in the body, which results in a dramatic increase in temperature, is an already well-established practice in the medical field[8]–[13]. Additionally, hydrophonic[14]–[16] and optical measuring methods[17]–[19] have been actively developed.

The idea of visualizing ultrasound by converting it into heat was proposed 50 years ago[20] and has been continuously studied ever since[21]–[24]. This method enables the rapid scanning of three-dimensional (3D) ultrasound fields with a simple setup[23].

However, this idea has not yet been applied to airborne ultrasound measurements, and only a few studies have investigated the conversion of airborne ultrasound into heat, such as in the acceleration of food drying[25], [26]. To the best of our knowledge, only two studies have measured the temperature change caused by the absorption of airborne ultrasound: one on ultrasonic ignition conducted by Simon et al.[27] and another on noncontact thermal display performed by our research group[28].

One reason for the lack of studies on airborne ultrasound is that high-power airborne ultrasound has received little attention, except for use in special engineering applications[29]. However, over the last decade, the demand for airborne ultrasound measurements has been growing in research fields such as mid-air haptics[30], [31] and acoustic levitation[32]–[34].

In this letter, we demonstrate that thermal images can be used to visualize an acoustic field of airborne ultrasound. The high-intensity airborne ultrasound produced by phased arrays causes a noticeable increase in temperature, and the sound pressure can be estimated based on the degree of temperature change. The acoustic field can be visualized without strong interference by using a thin mesh that allows most of the ultrasound waves to pass through. We also show the possibility to measure the ultrasound fields on the surfaces of various objects.

Experimental setup used in this study comprises a thermography camera and ultrasound phased arrays. The thermography camera (OPTPI 45ILTO29T090, Optris) had an optical resolution of 382 × 288 px, temperature resolution of 40 mK, and frame rate of 27 Hz. The ultrasound phased arrays can generate various acoustic field distributions by controlling the phase and amplitude of each transducer[35], [36]. To generate the acoustic field, we used four airborne ultrasound phased arrays consisting of 249 ultrasound transducers (T4010A1, Nippon Ceramic Co, Ltd.), as well as employed the same software architecture developed by Suzuki et al.[37]. In addition, we employed a nylon 6.6 mesh (PA-77μ, AS ONE). During the experiments, two types of acoustic fields were generated: one with a single focus and another consisting of a standing wave. The phased arrays were set in opposing positions to generate a standing wave but set in the same direction to generate a single focus.

Herein, we experimentally demonstrate the visualization of an airborne acoustic field using a temperature distribution. First, we confirmed that the thermal images on a mesh screen placed inside an acoustic field revealed the acoustic field in real time. Fig. 1 shows the actual thermal images captured during the experiment, from which one can identify features of the acoustic field such as the nodes and antinodes of the standing wave, and the side lobes around the focus. We refer to this measurement scheme as "Meshgraphy." Next, we validated that the thermal images can be used to visualize the acoustic field on the surface of an object that almost completely reflects ultrasound—in our case, a hand. Fig. 2 shows the resulting thermal images acquired during the experiment. We can observe the distribution of the acoustic field on the palm of the hand and the focal point moving across the fingertip with a resolution of less than the ~8.5 mm wavelength of the emitted 40 kHz ultrasound. These results demonstrate that the surface temperature of an object can be used to visualize the acoustic field generated by airborne ultrasound. We refer to this scheme as "Surfacegraphy" in this paper.

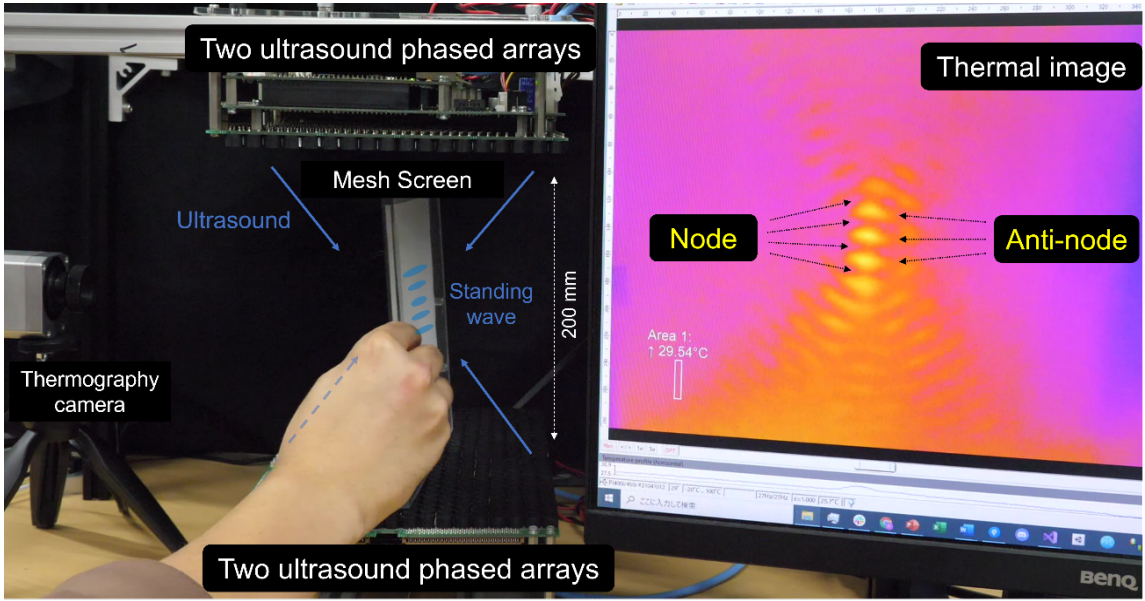

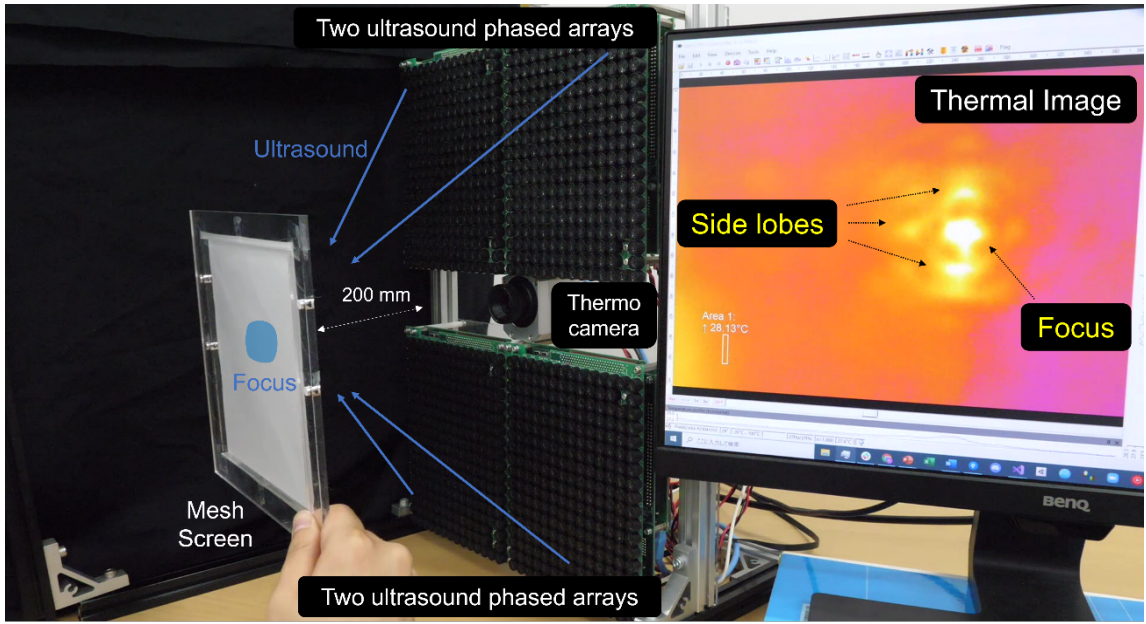

**Fig. 1 Visualization of the acoustic field on a mesh (Meshgraphy).** The image shows a mesh screen, four ultrasound phased arrays, a thermography camera, and one frame of thermal images on the display. **(a)** First, the ultrasound phased arrays generated an acoustic standing wave. Then, the mesh screen was manually placed inside the acoustic field. The frame describes the thermal distribution on the mesh screen. We can see that the frame can be used to visualize the features of the standing wave, particularly the nodes and antinodes. **(b)** The setup is the same as that in (a) except for the position of the infrared camera and the arrangement of the four ultrasound phased arrays, which were modified to generate an acoustic field with a single focus. We can see that the frame reveals the acoustic field, including its focal point and side lobes.

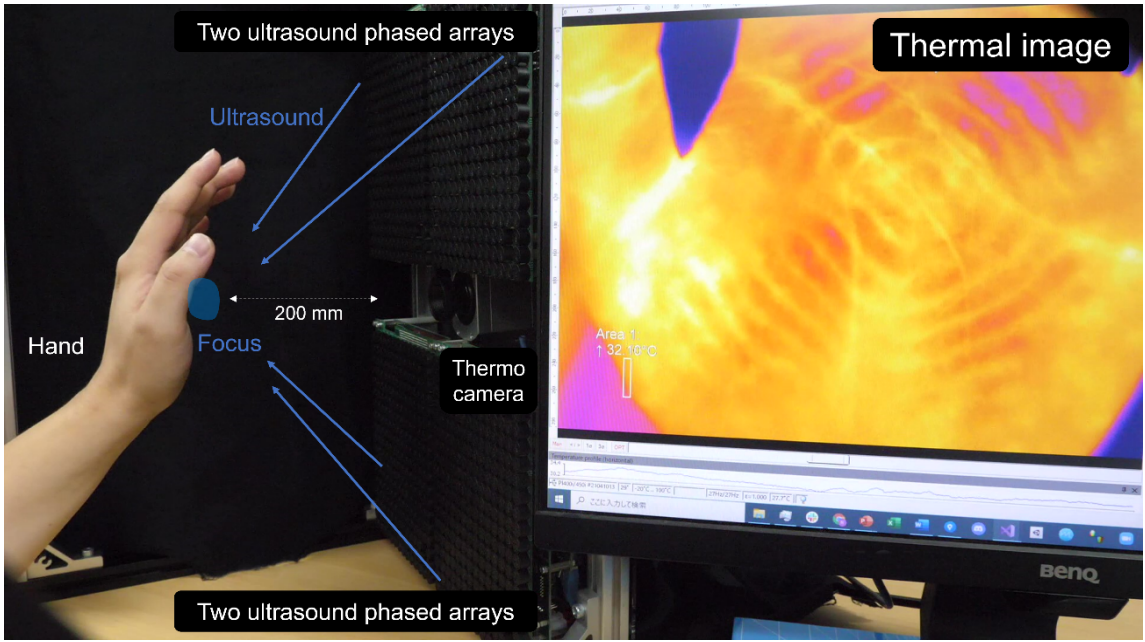

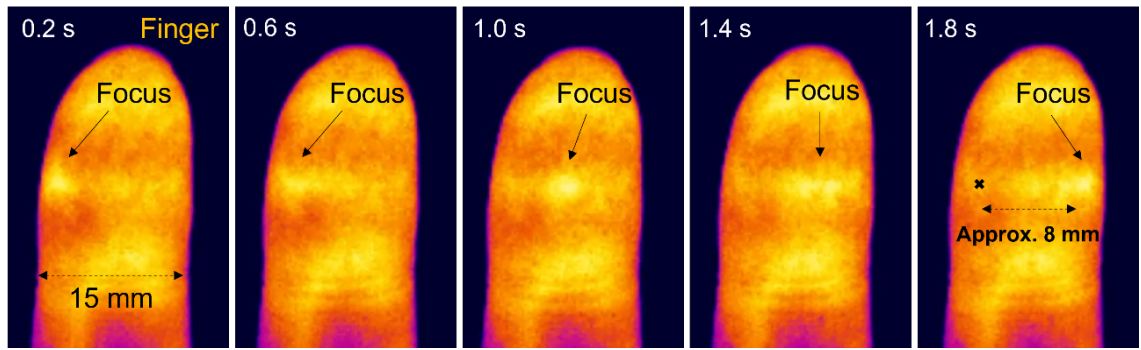

**Fig. 2 Visualization of the acoustic field on a reflective surface (Surfacegraphy). (a)** The image shows a hand, four ultrasound phased arrays, a thermography camera, and one frame of thermal images. First, the ultrasound phased arrays generated an acoustic field with a single focus. Then, the author placed the left hand in front of the acoustic field. The frame describes the thermal distribution on the hand. We can see a concentric circle pattern that visualizes the acoustic field on the palm. **(b)** The image sequence shows the thermal distribution on the fingertip obtained using the same environment as that in (a). The focus was moved in steps of 2 mm every 0.4 s (0.2 s irradiation and 0.2 s stop). We can confirm that the focus moved in 2 mm steps, resulting in a movement of ~8 mm.

Here, we identified the acoustic field aspect that is represented by the temperature distribution and the heat-generating mechanism. First, we checked the quantitative relationship between the sound pressure and the temperature increase ratio in Meshgraphy. We measured the temperature change at the location of the focus in the mesh surface, as well as the sound pressure at the focus using a standard microphone (Type 4138-A-015, Br€uel & Kjær), during single-focus generation with a single phased array module. In the experiment, we started the ultrasound output after 10 s and stopped it after 40 s. The sound pressure at the focus was 1146 Pa. The time-series of the mesh surface Temperature are plotted in Fig. 3 (a). Fig. 3 (b) shows that the focus temperature increased almost linearly by 0.8 °C after 1 s of ultrasound irradiation. In addition, we observed that the temperature reached an equilibrium state in ~4 s. This result confirms that a measurable temperature increase occurred at a high sound pressure of 1146 Pa. Moreover, the linear increase in temperature until equilibrium indicated that during this period, the cooling effect due to heat conduction through the mesh or thermal radiation to the air was negligible. During the experiment, the room temperature and humidity were 26.3 °C and 26.6%, respectively.

a

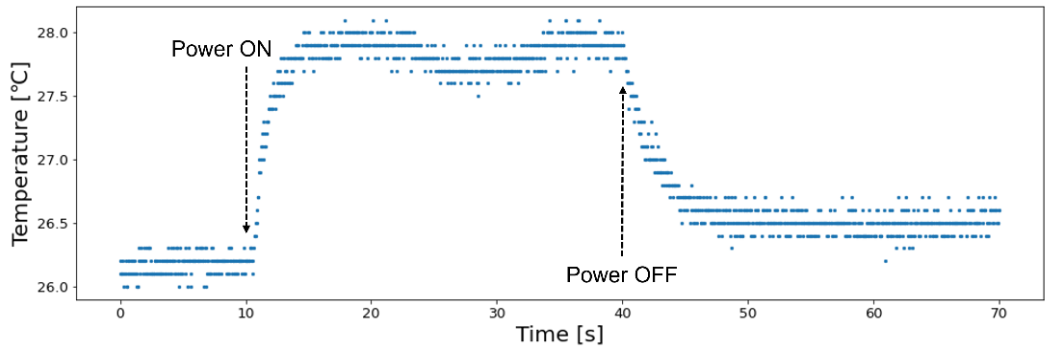

b

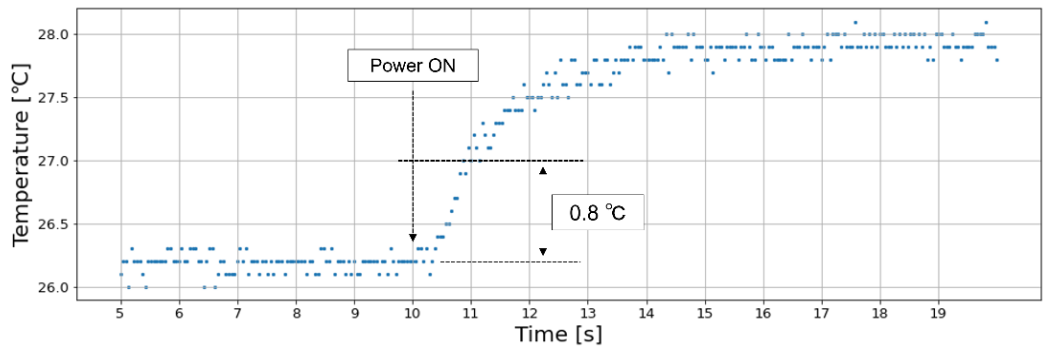

**Fig. 3 (a)** Time-series of the temperature at the focal point of the ultrasound field in Meshgraphy; the ultrasound irradiation started at t = 10 s and stopped at t = 40 s. **(b)** Enlarged image of the period between t = 5 s and 20 s, which shows a temperature increase of 0.8 °C in 1 s.

Next, to elucidate the mechanism behind the temperature rise, we investigated the relationship between sound pressure and temperature in a standing wave field. Two factors were considered to contribute to the temperature rise: heat transfer to the mesh from the surrounding air heated by the ultrasound waves[38] and viscous heating due to the interaction between the mesh and air[39]. In the antinode of the standing wave, the particle velocity was almost zero, and there was little heat generation due to viscosity. Thus, we could assume that if the temperature increased in the antinode, the contribution from heat transfer was significant. Conversely, if the temperature increased at the node of the standing wave where the particle velocity is maximum, we may assume that the contribution from viscous heating was significant. In this experiment, we used two phased arrays that were facing each other to generate a standing wave. Fig. 4 (a) shows the temperature distribution at the surface of the mesh placed within the standing wave 0.5 s after the start of the ultrasound irradiation: x and z represent the spatial positions. Subsequently, the sound pressure at (x = 0, -30 < z < 30) in Fig. 4 (a) was scanned with a microphone at 0.1 mm intervals. Because the microphone could only measure sound pressures of up to 5000 Pa, the ultrasonic power was reduced to 3 % for accurate measurement. We plotted the position $z$ on the horizontal axis and the temperature $T$ [°C] and sound pressure $P$ [Pa] on the vertical axis in Fig. 4 (b). This figure demonstrates that the temperature increases at the point where the sound pressure is high, which corresponds to the antinode of the standing wave. This indicates that heat transfer to the mesh from the surrounding air heated by ultrasound waves is the major contributing factor to the increase in the mesh temperature. The room temperature and humidity were 21.6 °C and 25.5 %, respectively.

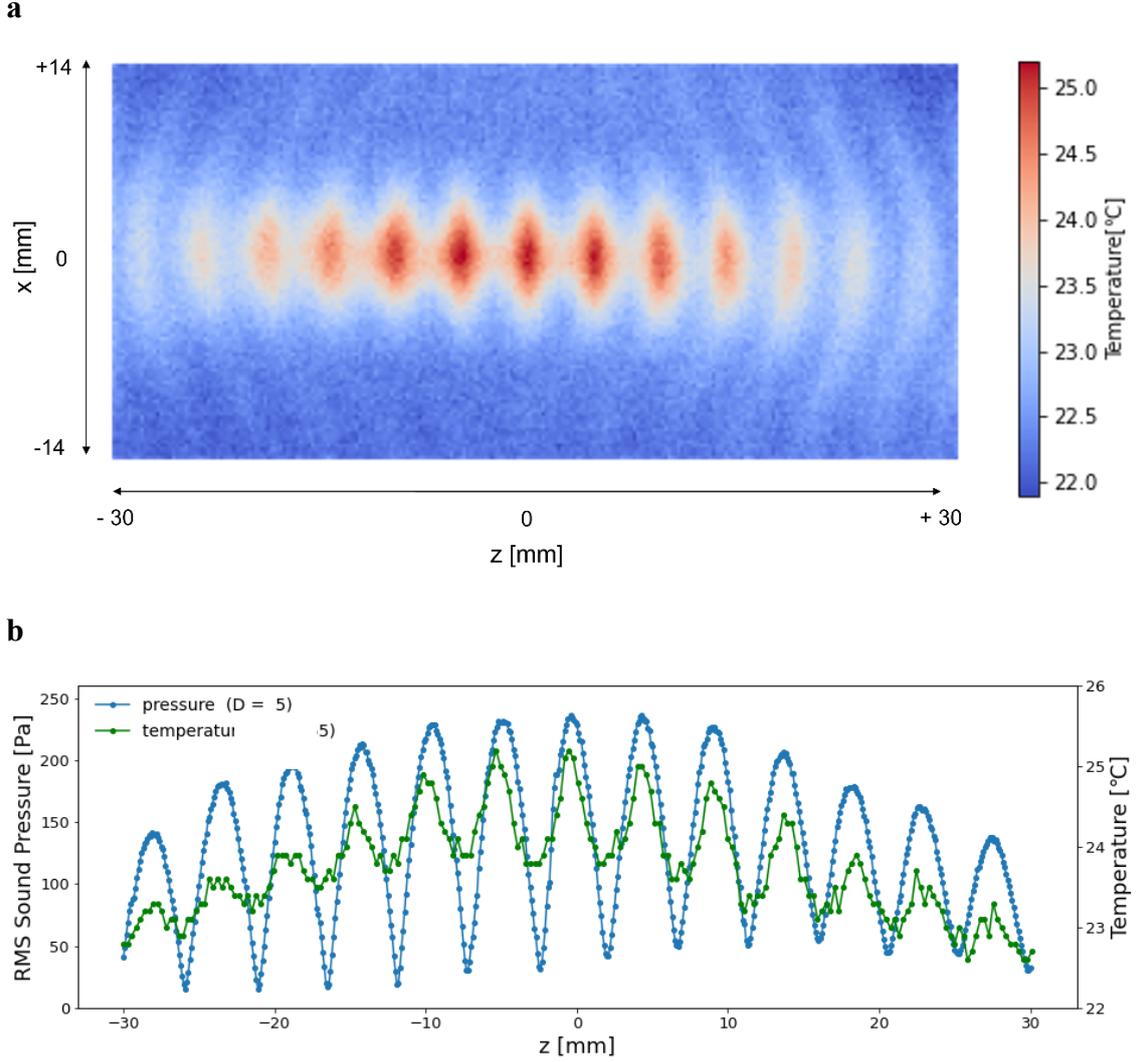

**Fig. 4 (a)** Temperature distribution on a mesh placed within a standing wave. x and z represent spatial positions. **(b)** The horizontal axis is the z position, and the vertical axes are temperature and sound pressure. This figure demonstrates that the temperature increases significantly at the antinode of the standing wave, where the sound pressure is high.

Finally, we compared the sound pressure estimated from temperature changes and that measured with the microphone. According to previous studies[38], [40], [41], the temperature of the medium $T$ and the acoustic intensity $I$ have the following relationship:

$$\rho C \frac{\partial T}{\partial t} = \nabla \cdot (k \nabla T) + 2\alpha I \qquad (1)$$

where $\rho$, $C$, $k$, and $\alpha$ denote the medium density, specific heat, thermal conductivity, and attenuation coefficient, respectively. If we consider that the temperature on the nylon mesh surface, $T_n$, and the temperature of the air medium, $T$, are the same, and assume that the temperature gradient $\nabla T$ in the focal diameter is negligible, Equation (1) becomes

$$\rho C \frac{\partial T_n}{\partial t} = 2\alpha I \qquad (2)$$

Furthermore, the acoustic intensity $I$ can be expressed as

$$I = pu = \frac{p^2}{\rho c}$$

where $p$ is the RMS of the sound pressure, and $c$ is the speed of sound. Considering this, Equation (2) becomes:

$$p = \sqrt{\frac{\rho^2 cC}{2\alpha}} \times \sqrt{\frac{\partial T_n}{\partial t}} \qquad (3)$$

According to a previous study [35], the attenuation coefficient $\alpha$ is $1.15 \times 10^{-1}$ Np/m. If we consider that the temperature change ratio $\partial T_n/\partial t$ is 0.8 K/s from Fig. 3, $\rho$ is 1.29 kg/m$^3$, $c$ is 340 m/s, and $C$ is 1006 J/kg·K, we can obtain a $p$ value of 1407 Pa using Equation (3). The sound pressure measured by the microphone was 1146 Pa. Thus, the theoretical predictions agreed well with the measured results. Note that the calculations were performed by ignoring the heat capacity of the nylon mesh. The heat capacity of the nylon mesh was not sufficiently small to be ignored completely, causing the measured temperature change rate $\partial T_n/\partial t$ to be smaller than the theoretical value, but it can be considered to have an insignificant effect on this result.

In summary, we experimentally demonstrated that thermal images can be used to visualize acoustic fields in real-time. We visualized an airborne ultrasound field generated by phased arrays with an output of more than 1000 Pa, using a thermography camera with a standard temperature resolution of 40 mK (Meshgraphy). We also visualized an airborne ultrasound field on the surface of an object that almost completely reflects ultrasound (Surfacegraphy). Based on the observation of temperature increase of the pressure antinode of the standing wave, we concluded that this phenomenon in Meshgraphy is caused by heat transfer to the mesh from the surrounding air heated by ultrasound waves. Meanwhile, the quantitative mechanism of Surfacegraphy has been left as future work. The rapid and detailed visualization of acoustic fields based on our method paves the way for research on the use of strong airborne ultrasound. At the least, our findings may play a significant role in elucidating tactile perception mechanisms using ultrasonic tactile stimulations that require high precision in the presentation position and force. Such high-intensity airborne ultrasound also provides a general scientific tool as a non-contact mechanical actuator to manipulate and evaluate elastic and fragile objects, including cells.

**Acknowledgements**
This work was supported in part by JST CREST JPMJCR18A2 and JSPS KAKENHI Grant Number JP20K19841.

**Conflict of Interest**
Authors declare no competing interests.